\documentclass{aastex}
\usepackage{spr-astr-addons}
\usepackage{url}
\usepackage{lscape, morefloats, graphicx, float} 

\RequirePackage{color}

\begin{document}

\title{CCD ${\it UBVRI}$ photometry of NGC 6811}
\slugcomment{Not to appear in Nonlearned J., 45.}
\shorttitle{CCD ${\it UBVRI}$ photometry of NGC 6811}
\shortauthors{T. Yontan, S. Bilir, Z. F. Bostanc\i, T. Ak, S. Karaali, T. G\"uver, S. Ak, \c S. Duran, \and E. Paunzen}

\author{T. Yontan \altaffilmark{1}}
\altaffiltext{1}{Istanbul University, Graduate School of Science and Engineering, 
Department of Astronomy and Space Sciences, 34116, Beyaz\i t, Istanbul, Turkey\\
\email{talar.yontan@gmail.com}}

\author{S. Bilir \altaffilmark{2}}
\altaffiltext{1}{Istanbul University, Faculty of Science, Department 
of Astronomy and Space Sciences, 34119 University, Istanbul, Turkey\\}
\and
\author{Z. F. Bostanc\i\altaffilmark{2}} 
\altaffiltext{2}{Istanbul University, Faculty of Science, Department 
of Astronomy and Space Sciences, 34119 University, Istanbul, Turkey\\}
\author{T. Ak\altaffilmark{2}} 
\altaffiltext{2}{Istanbul University, Faculty of Science, Department 
of Astronomy and Space Sciences, 34119 University, Istanbul, Turkey\\}
\and
\author{S. Karaali\altaffilmark{2}} 
\altaffiltext{2}{Istanbul University, Faculty of Science, Department 
of Astronomy and Space Sciences, 34119 University, Istanbul, Turkey\\}
\and
\author{T. G\"uver\altaffilmark{2}} 
\altaffiltext{2}{Istanbul University, Faculty of Science, Department 
of Astronomy and Space Sciences, 34119 University, Istanbul, Turkey\\}
\and
\author{S. Ak\altaffilmark{2}} 
\altaffiltext{2}{Istanbul University, Faculty of Science, Department 
of Astronomy and Space Sciences, 34119 University, Istanbul, Turkey\\}
\and
\author{\c S. Duran\altaffilmark{1}} 
{\altaffiltext{1}{Istanbul University, Graduate School of Science and Engineering, 
Department of Astronomy and Space Sciences, 34116, Beyaz\i t, Istanbul, Turkey\\}
\and
\author{E. Paunzen\altaffilmark{3}} 
{\altaffiltext{3}{Department of Theoretical Physics and Astrophysics, 
Masaryk University, Kotl\'a\u rsk\'a 2, 611 37 Brno, Czech Republic\\}

\begin{abstract}
We present the results of CCD $UBVRI$ observations of the open cluster NGC 6811 obtained on 18th July 2012 with the 1 m telescope at the T\"UB\.ITAK National Observatory (TUG). Using these photometric results, we determine the structural and astrophysical parameters of the cluster. The mean photometric uncertainties are better than 0.02 mag in the $V$ magnitude and $B-V$, $V-R$, and $V-I$ colour indices to about 0.03 mag for $U-B$ among stars brighter than magnitude $V=18$. Cluster member stars were separated from the field stars using the {\it Galaxia} model of \citet{Sharma11} together with other techniques. The core radius of the cluster is found to be $r_{c}=3.60$ arcmin. The astrophysical parameters were determined simultaneously via Bayesian statistics using the colour-magnitude diagrams $V$ versus $B-V$, $V$ versus $V-I$, $V$ versus $V-R$, and $V$ versus $R-I$ of the cluster. The resulting most likely parameters were further confirmed using independent methods, removing any possible degeneracies. The colour excess, distance modulus, metallicity and the age of the cluster are determined simultaneously as $E(B-V)=0.05\pm0.01$ mag, $\mu=10.06\pm0.08$ mag, $[M/H]=-0.10\pm0.01$ dex and $t = 1.00\pm0.05$ Gyr, respectively. Distances of five red clump stars which were found to be members of the cluster further confirm our distance estimation.
\end{abstract}


\section{INTRODUCTION}
Galactic clusters are distributed along the Galactic disc, with scaleheights consistent with thin disc population. This distribution makes them ideal tracers of the Galactic disc, allowing for a better understanding of the Galactic structure, chemical composition, stellar population and star formation processes in the Galaxy. Thus, the structure and astrophysical parameters of individual clusters such as age, metal abundance, distance and reddening has to be determined through observations of these systems. The most practical and therefore most often used observational techniques are based on utilising colour-magnitude and two-colour diagrams observed in the optical bands.

The intermediate-age open cluster NGC 6811 ($l=79^{\circ}.210$, $b=+12^{\circ}.015$) is located in the Kepler field \citep{Borucki11}. Historically, this cluster has seldomly been studied through only a limited number of photographic and photometric observations. First $UBV$ photographic measurements of the cluster were performed by  \citet{Lindoff71}, finding that the age of the cluster is $5\times 10^8$ yr and located at a distance of 1100 pc using 377 stars within the line-of-sight of the cluster. \citet{Barkhatova78} estimated the age and distance as $8\times10^8$ yr and 1150 pc, respectively, with a larger sample of stars (2000 stars) brighter than $V=15.2$ magnitude. \citet{Sanders71} identified 97 members of the cluster by investigating the proper motions of 296 stars brighter than $V=14$ magnitude in the field of NGC 6811. \citet{Mermilliod90} estimated the radial velocity of the cluster as $V_R=7.28$ km~s$^{-1}$ by means of the data of three cluster members. \citet{Glushkova99} studied the cluster photometrically and spectroscopically. They measured the distance and age of the cluster as 1040 pc and  $7\times10^8$ yr, respectively. The most recent observations of NGC 6811 were reported by \citet{Janes13} using CCD $UBVRI$ photometric data obtained from different telescopes. Following a Bayesian approach they measured colour excess, distance modulus, metal abundance and the age of the cluster as $E(B-V)=0.074 \pm 0.024$ mag, $(m-M)_V=10.22 \pm 0.18$ mag, $Z= 0.012 \pm 0.004$ ($[M/H]=-0.19 \pm 0.03$ dex) and $t=1.00 \pm 0.17$ Gyr, respectively. 
  
Here, we provide colour excesses, distance modulus and metal abundance of the cluster following  different and independent approaches. On the one hand, we followed Bayesian methods, where the individual parameters are determined simultaneously. On the other hand, we used methods that involve comparing theoretical isochrones with the observed two-colour and colour-magnitude diagrams of the cluster members. The main difference between these procedures is that the parameters resulting from the Bayesian statistics may be affected from degeneracy. We aim to compare the astrophysical parameters of NGC 6811 obtained through a Bayesian approach and independent -traditional- methods. Although the degeneracies are possible in simultaneous solution of the parameters through the Bayesian approach, they are minimized as the $U-B$ colour and the possible red giants help to separate reddening and metallicity.

The layout of the paper is as follows: Section 2 presents the observations and photometric reduction of the open cluster NGC 6811. Section 3 describes colour-magnitude diagrams, and structural parameters of the cluster. In Section 4, we derive the astrophysical parameters of the cluster using a Bayesian approach. Section 5 is devoted to the independent methods used to determine the astrophysical parameters. Finally, in Section 6 we present a discussion and a conclusion of our study.
  
\section{OBSERVATIONS}
CCD $UBVRI$ observations of NGC 6811 were carried out on 18th July 2012 using the 1m Ritchey-Chr\'etien telescope (T100) located at the T\"UB\.ITAK National Observatory (TUG)\footnote{www.tug.tubitak.gov.tr} in Bak{\i}rl{\i}tepe, Antalya/Turkey. The mounted camera was an SI~1100 CCD camera (back illuminated, 4k$\times$4k pixels). The pixel scale is $0.''31$ pixel$^{-1}$, resulting in a total field of view of about $21.'5 \times 21.'5$. The readout noise of the system is 4.19 e$^{-}$, and the gain is 0.55~e$^{-}$/ADU. The cluster field was observed using short and long exposure times in each filter to be able to cover the widest possible flux range, without being affected by photon pile-up. Log of observations is given in Table 1. The night was moderately photometric with a mean seeing of $1''.5$. 

We used Image Reduction and Analysis Facility (IRAF)\footnote{IRAF is distributed by the National Optical Astronomy Observatories} routines for pre-reduction processes and transforming the pixel coordinates of the objects identified in frames to equatorial coordinates. We observed several standard stars selected from \citet{Landolt09} during the night for the determination of atmospheric extinction and transformation coefficients for the observing system. We determined the instrumental magnitudes of the standard stars, utilizing IRAF software packages with aperture photometry. For the field of the cluster, we used Source Extractor (SExtractor)\footnote{SExtractor: Software for source extraction} and isophotal photometry \citep{Bertin96}. In order to perform aperture corrections for the instrumental magnitudes obtained from isophotal photometry, we also performed aperture photometry of a number of well separated stars in the field. Then, the resulting aperture corrections were applied to the instrumental magnitudes of the stars in the field. Finally, we used the following equations \citep{Janes11,Janes13} to transform the instrumental magnitudes and colours of stars to the standard photometric system:

for $V$, $B-V$ and $U-B$;
\begin{eqnarray}
B-V = \frac{(b-v)-(k_b-k_v)X_{bv}-(C_b-C_{bv})}{\alpha_b+k'_bX_b-\alpha_{bv}} \\
V = v - \alpha_{bv}(B-V)-k_vX _v- C_{bv}\\
U-B = \frac{(u-b)-(1-\alpha_b-k'_bX_b)(B-V)}{\alpha_{ub}+k'_uX_u} \\ \nonumber
-\frac{(k_u-k_b)X_{ub}-(C_{ub}-C_b)}{\alpha_{ub}+k'_uX_u} 
\end{eqnarray}

for $V$ and $V-R$;
\begin{eqnarray}
V-R = \frac{(v-r)-(k_v-k_r)X_{vr}-(C_{vr}-C_r)}{\alpha_{vr}-\alpha_r} \\
V = v - \alpha_{vr}(V-R)-k_vX_v - C_{vr} 
\end{eqnarray}

for $V$ and $V-I$;
\begin{eqnarray}
V-R = \frac{(v-i)-(k_v-k_i)X_{vi}-(C_{vi}-C_i)}{\alpha_{vi}-\alpha_i} \\
V = v - \alpha_{vi}(V-I)-k_vX_v - C_{vi}, 
\end{eqnarray}
where $U, B, V, R$ and $I$ are defined as magnitudes in the standard photometric system, $u$, $b$, $v$, $r$ and $i$ the instrumental magnitudes and $X$ the airmass. $k$ and $k'$ are primary and secondary extinction coefficients while $\alpha$ and $C$ are transformation coefficients to the standard system. Applying multiple linear fits to the instrumental magnitudes of the standard stars, we obtained the photometric extinction and transformation coefficients for that particular night. The resulting values are given in Table 2. 

\section{DATA ANALYSIS}
\subsection{Identification of Stars and Error Analyses}

Using the SExtractor, we detected 1605 objects in the field of NGC 6811 and constructed a catalogue. In order to identify non-stellar objects, most likely galaxies, in our catalog, we used the the stellarity index (SI) provided by SExtractor. The apparent $V$ magnitude versus stellarity diagram of 1605 objects is plotted in Fig. 1. The objects with stellarity index smaller than 0.8 are assumed to be non-stellar sources and therefore excluded from further analysis \citep{Andreuzzi02,Karaali04}. The resulting catalog has 1591 objects that are classified as stars. The final photometric catalogue containing the colours and magnitudes of individual stars are tabulated in Table 3. 

The mean errors of the measurements in the $V$ band and $U-B$, $B-V$, $V-R$, $V-I$ and $R-I$ colours are given in Table 4 and Fig. 2 as a function of the apparent $V$ magnitude. The discontinuity of the errors at $V=13.8$ mag in Fig. 2 originates from the use of two different exposure times. The errors are relatively small for stars with $V<18$ mag, while they increase exponentially towards fainter magnitudes. The largest errors for a given $V$ magnitude belong to $U-B$ colour. For stars brighter than $V$ magnitude 15 and stars with $V$ magnitudes between 15 and 18, the mean photometric errors in the $V$, $B-V$, $V-R$, and $V-I$ indices are smaller than 0.004 and 0.022 mag, respectively. The errors in the $U-B$ colour indices are smaller than 0.004 and 0.034 mag for the same $V$ magnitude ranges. 

As stated above, NGC 6811 has been the subject of many studies. The most recent study for this cluster is carried out by \cite{Janes13} who used the same photometry, i.e. $UBVRI$, in their investigation, as it is located in the {\it Kepler} spacecraft's field of view \citep{Stello11, Meibom11, Hekker11, Corsaro12}. We cross-matched our catalogue with the catalogue of \cite{Janes13} to compare the photometric measurements. We identified 653 stars that are detected in both catalogues. The results of these comparisons are given in Fig. 3. The magnitudes and colours on the abscissa refer to the ones in our study, while the magnitude and colour differences on the ordinate show the differences between the two catalogues. The means and the standard deviations of the resulting differences are small except for $\Delta U$ and $\Delta (U-B)$, i.e. $\sigma_{U}=0.079$ and $\sigma_{U-B}=0.067$ mag. The agreement is much better in the $R$ and $I$ bands and their related colours.

\subsection{Colour-Magnitude Diagrams}

The most effective tool for the determination of the parameters for a cluster is its colour-magnitude diagrams. We used  four colour-magnitude diagrams, i.e. 
$V$ versus $U-B$, $V$ versus $B-V$, $V$ versus $V-R$ and $V$ versus $R-I$ for the stars in Table 3, for this purpose. A glance to the colour-magnitude diagrams in Fig. 4 reveals that our cluster is rather sparse. Although most of the stars brighter than $V=16$ mag are lying along a sequence, resembling the main-sequence of an open  cluster, the vertical distribution of the stars fainter than $V=16$ mag imply that most of these stars are field stars. 

In all panels, among the bright and red stars half a dozen of them occupy the region where red clump (RC) stars lie, i.e. $0.7\leq (B-V)_0\leq 1.2$ mag \citep{Bilir13a}. These stars are important because they have been used as standard candles \citep[i.e.][]{Paczynski98,Cabrera2005}. These RC stars can be used to confirm the distance of a cluster found from other methods. The position of another small group of stars, bright but blue, implies that the turn-off point of the cluster lie within $11.5<V<12$ mag. The mean $B-V$ colour of these stars is about 0.2 mag. 

As detailed below, in the identification of the cluster members we followed a two-step approach; first we made certain selections using theoretically calculated isochrones, then we used the {\it Galaxia} model \citep{Sharma11} to separate the field stars from the cluster members. 

We assumed that 387 stars with $10\leq V\leq18$ mag are likely evolved and main-sequence members of the cluster and included in our further analysis due to the sensitivity of our observations (Fig. 5a). Below $V=12$ mag, to identify the likely main-sequence members of the cluster we applied the following selections. We fitted the zero age main-sequence (ZAMS) Padova isochrone with solar metallicity to 357 stars within an apparent magnitude interval $12\leq V\leq18$ and moved it to brighter $V$ magnitudes by 0.75 mag to cover the binary stars, as well, resulting in a band like region in the colour-magnitude diagram (see Fig. 5a).  

We used the {\it Galaxia} model of \cite{Sharma11} and estimated the $V$ magnitudes and $B-V$ colours of stars, which are expected to lie in the direction to the cluster NGC 6811 (Fig. 5b). We estimated the colours and magnitudes of these synthetic stars in different Galactic populations using the input parameters of the {\it Galaxia} model for the Galactic coordinates ($l=79^{\circ}.210, b=+12^{\circ}.015$), spatial size (0.1136 square-degree), colour ($0<B-V<1.5$ mag) and the range of apparent magnitude ($10\leq V\leq 18$) of the cluster. Then, we omitted those which fall into the categories of bulge, thick disc and halo stars plus the thin-disc stars older than 5 Gyr. Note that a small number of older stars could fall into the relevant region of the colour-magnitude diagram. Resulting sample contains 446 synthetic stars in the line-of-sight of NGC 6811. As a final restriction, we selected only those which are located in the same band on the colour-magnitude diagram Fig. 5b. The final number of stars which satisfy all of these conditions is 133. The next step in the selection is to subtract randomly selected 133 stars within the  apparent magnitude interval of $10\leq V\leq 18$ with 1 mag steps in the colour-magnitude diagram of the cluster as shown in Fig. 5a. Thus the resulting number of cluster stars within the apparent magnitude interval $10\leq V\leq 18$, with zero age and solar metallicity is 254. The $V$ versus $B-V$ diagram of the remaining 254 stars is shown in Fig. 5c.

\subsection{Cluster Radius and Radial Stellar Surface Density}
The stellar density profile of the cluster NGC 6811 has been estimated in two different ways. In the first case, we used only the main-sequence stars of the cluster given in Fig. 5c, while in the second case both categories, i.e. the cluster stars and the field main-sequence stars within the same apparent $V$ magnitude interval, with the same age and metallicity as cluster main-sequence stars (Fig. 5a) have been introduced into our calculations. The stellar density values for the main-sequence sample of the cluster have been evaluated in 2 arcminute steps, while for the second sample 1 arcminute step was sufficient for reliable  density evaluations.

Stellar density profiles of the cluster are plotted in Fig. 6. We fitted each of them the \cite{King62} model defined as: 
\begin{equation} 
\rho(r)=f_{bg}+\frac{f_{0}}{1+(r/r_{c})^{2}},
\end{equation} 
where $r$ is the radius of the cluster centered at the Galactic coordinates ($l$, $b$)=($79^{\circ}.210, +12^{\circ}.015$). $f_{bg}$, $f_0$ and $r_c$ denote the background stellar density, the central stellar density and the core radius of the cluster, respectively. The core radii estimated for two stellar samples are compatible within the errors, i.e. $3.600\pm0.563$ and $3.860\pm0.275$ arcmin for the cluster stars and cluster plus field stars, respectively. However, the central densities for the two stellar samples are different than each other, as expected. The $f_0$ central stellar densities are $0.921\pm0.115$ and $1.566\pm0.085$ stars/arcmin$^2$. 

\section{DETERMINATION OF THE ASTROPHYSICAL PARAMETERS OF NGC 6811 USING A BAYESIAN APPROACH}

We estimated the age of the cluster NGC 6811 using the procedure of \cite{Jorgensen05} as explained in the following. We quote \citeauthor{Jorgensen05}'s paper for details, while a shorter description can be found in \citet{Duran13}. The procedure we used is based on the posterior joint probability defined as follows:
\begin{equation} 
f(\tau,\zeta,m) \propto f_0(\tau,\zeta,m)~ L(\tau,\zeta,m) 
\end{equation} 
where $f_o$ and and $L$ are the prior probability density of the parameters and the likelihood function, respectively. The parameters $\tau$, $\zeta$, $m$ are the age, initial metallicity and initial mass, respectively. The probability density function is defined such that $f(\tau,\zeta,m)\mbox{d}\tau\mbox{d}\zeta\mbox{d}m$ is the fraction of stars with ages between $\tau$ and $\tau+\mbox{d}\tau$, 
metallicities between $\zeta$ and $\zeta+\mbox{d}\zeta$, and initial masses between $m$ and $m+dm$. The constant of proportionality in Eq. (9) 
must be chosen to make $\int\!\int\!\int f(\tau,\zeta,m)\mbox{d}\tau\mbox{d}\zeta\mbox{d}m=1$.    

The likelihood function ($L$) is equal to the probability of getting the observed data $q$(colour, apparent magnitude, metallicity) for given parameters $p(\tau,\zeta,m)$. Then, the likelihood function is  
\begin{equation} 
L(\tau,\zeta,m) = \left( \prod_{i=1}^n \frac{1}{(2\pi)^{1/2}\sigma_i} \right) \times \exp(-\chi^2/2), 
\end{equation}
where
\begin{equation}
\chi^2 = \sum_{i=1}^n \left(\frac{q_i^{\rm obs}-q_i(\tau,\zeta,m)}{\sigma_i}\right)^2,
\end{equation} 
and where $\sigma_i$ is the standard error. A maximum-likelihood calculation of the stellar parameters $(\tau,\zeta,m)$ may be obtained by finding the maximum of this function, which is equivalent to minimizing $\chi^2$ in the case of Gaussian errors.

The prior density of the model parameters in Eq. (12) can be written as
\begin{equation}
f_0(\tau,\zeta,m) = \psi(\tau)\phi(\zeta\vert\tau)\xi(m\vert\zeta,\tau),
\end{equation} 
where $\psi(\tau)$ is the a priori star formation rate history, $\phi(\zeta\vert\tau)$ the metallicity distribution as a function of age, 
and $\xi(m\vert\zeta,\tau)$ the a priori initial mass function (IMF) as a function of metallicity and age. In our study we adopted the metallicity distribution as a flat distribution, and IMF the following power-law :
\begin{equation}
\xi(m) \propto m^{-\alpha},
\end{equation} 
with $\alpha=2.7$. These constraints which are the same as in \citet{Jorgensen05} and \citet{Duran13} provide simplifications in calculations. If we insert Eq. (12) into Eq. (9) and integrate with respect to $m$ and $\xi$ we obtain the following form of the posterior probability density function of $\tau$:
\begin{equation}
f(\tau) \propto \psi(\tau)G(\tau),
\end{equation}
where
\begin{equation}
G(\tau) \propto \int\!\int L(\tau,\zeta,m)\xi(m)~\mbox{d}m~\mbox{d}\zeta.
\end{equation}
We normalize Eq. (15) such that $G(\tau)=1$ at its maximum. \citet{Jorgensen05} interpreted $G(\tau)$ as the relative likelihood of $\tau$ after eliminating $m$ and $\zeta$. 

Following \citet{Jorgensen05}, we evaluated Eq. (15) for each age value ($\tau_i$) as a double sum along a set of isochrones at the required age that are equidistant in metallicity ($\zeta_k$). In practice, we used pre-computed isochrones for a step size of 0.05 dex in $\zeta$, and considered only those within $\pm3.5\sigma_{[M/H]}$ of the observed metallicity. Let $m_{jkl}$ be the initial-mass values along each isochrone ($\tau_j$, $\zeta_k$); then 
\begin{equation}
G(\tau_j) \propto \sum_k \sum_\ell L(\tau_j,\zeta_k,m_{jk\ell}) \xi(m_{jk\ell})(m_{jk\ell+1}-m_{jk\ell-1}).
\end{equation}
Age corresponding to the mode of the relative posterior probability $G(\tau)$ is adopted as the age of the star in question.  

In the case of the cluster NGC 6811, we adopted the Padova synthetic stellar library \citep{Marigo08}. We used the isochrones with 0-3 Gyr in steps of 50 Myr and the metallicities $-0.3\leq[M/H]\leq+0.2$ dex in steps of 0.05 dex. For the distance modulus we used the range as $9\leq \mu \leq11$ mag in steps of 0.01 mag. The observed data consist of the $B-V$, $V-R$, $V-I$ and $R-I$ colours and $V$ apparent magnitude. Because we have four sets of colour-magnitude diagrams for the cluster stars (Fig. 7), we estimated four sets of ages for the stellar sample. The distribution of the $G(\tau)$ parameter in terms of age for each set is shown in Fig. 8. Each distribution has a relatively high peak, as expected for the stars of a cluster. The mode of the distribution obtained from each colour-magnitude diagram (Table 5) corresponds to the age of the cluster. The final age of NGC 6811 is determined as $t=1.00\pm0.05$ Gyr, by averaging these individual results.

In each step of age ($\tau_j$) estimation for a given colour-magnitude diagram, we used a colour excess for de-reddening the corresponding distance modulus and colour indices, in addition to an amount of metallicity. The colour excesses $E(B-V)$, $E(V-R)$, $E(V-I)$ and $E(R-I)$ are selected from an interval of [0, 0.2] mag in steps of 0.01 mag. The colour excesses versus the corresponding apparent distance modulus and metallicity are plotted in Fig. 9. The most probable colour excess, distance modulus and metallicity for each colour-magnitude diagram are also indicated in the figure and the results are given in Table 5. The weighted mean of the four distance moduli and metallicities, i.e. $\mu=10.06\pm 0.08$ mag and $[M/H]=-0.10\pm0.01$ dex, are adopted as the distance modulus and the metallicity of the cluster NGC 6811. Resulting isochrones with the most likely parameter values are shown in Fig. 10 for each colour-magnitude diagram together with the uncertainties.

\section{INDEPENDENT CONFIRMATION OF THE ASTROPHYSICAL PARAMETERS}

In Section 4, we determined the astrophysical parameters of the cluster simultaneously using a Bayesian approach. However, in reality the distance, reddening, and the metallicity all affect the observed colour-magnitude diagrams of a cluster in similar ways. In order to break the degeneracies between these parameters and obtain the most reliable values, independent measurements are essential. Below, we use a number of well-known and independent methods to obtain astrophysical parameters of NGC 6811 and assess the results of the Bayesian approach.

\subsection{Reddening and Distance Modulus Estimation} 
We used main-sequence stars of the cluster within the magnitude range $12\leq V \leq 18$ for the determination of the reddening in five colours: $E(U-B)$, $E(B-V)$, $E(V-R)$, $E(V-I)$, and $E(R-I)$, comparing their positions in the two-colour diagrams with three standard main-sequences, as shown in Fig. 11. The first standard main-sequence \citep{Sung13} is constructed using the  observational data of the solar metallicity main-sequence stars for the $U-B$ versus $B-V$ colours, while the other two are taken from the Padova synthetic stellar library \citep{Marigo08} calculated for the solar metallicity and zero age stars in the colours $V-I$ versus $V-R$ and $R-I$ versus $V-R$, respectively. 

We adopted an amount of $E(B-V)$ colour excess values within the range $0-0.2$ mag in steps of 0.001 mag and used them in the following equations of \cite{Cardelli89} to evaluate the $E(U-B)$, $E(V-R)$, and $E(V-I)$ colour excesses:

{\small
\begin{eqnarray}
E(U-B)=E(B-V)\times [0.72+0.05\times E(B-V)],\\
E(V-R)=0.65\times E(B-V),\\
E(V-I)=1.25\times E(B-V).
\end{eqnarray}}

The positions of the cluster stars have been compared to the standard main sequence in question, and the most likely colour excesses have been determined by minimum $\chi^2$ analyses. The final reddened and de-reddened main-sequence curves fitted to the cluster stars are shown in Fig. 11. The contours of $\chi^2$ distributions are also given in the left corner of each diagram. The colour excesses estimated in this way are as follows: $E(U-B)=0.033\pm0.011$, $E(B-V)=0.046\pm0.012$, $E(V-R)=0.053\pm 0.010$, $E(V-I)=0.102\pm0.018$, and $E(R-I)=0.046\pm0.010$ mag.  

The distance modulus of the cluster NGC 6811 has been determined by fitting the colour-magnitude diagrams of the main-sequence stars with $12\leq V\leq18$ mag to four standard colour-magnitude diagrams (Fig. 7). For the $V$ versus $B-V$ diagram, we used the ZAMS given by \cite{Sung13}, while we adopted the corresponding ZAMS taken from the Padova synthetic stellar library \citep{Marigo08} for the $V$ versus $V-R$, $V$ versus $V-I$, and $V$ versus $R-I$. The ZAMS's are reddened by an amount of colour excesses found above and fitted to the cluster stars for different distance moduli values in the 9-11 mag range with a step size of 0.01 mag. The minimum $\chi^2$ statistics is used to obtain the most probable distance moduli for each colour-magnitude diagram. The resulting distance moduli are as follows: $\mu_{B-V}=10.22\pm 0.15$, $\mu_{V-R}=10.26\pm 0.25$, $\mu_{V-I}=10.22\pm 0.25$, and $\mu_{R-I}=10.32\pm 0.46$ mag. The final distance modulus of the cluster is the weighted mean of the distance moduli claimed for the four colour-magnitude diagrams, i.e. $\mu=10.15\pm 0.11$ mag. For the estimation of the errors, we adopted the procedure of \cite{Phleps00}, i.e. by adding 1 to the $\chi^2$ value.    

\subsection{Distance via the Red Clump Stars}
Distance estimation given above for the cluster can be confirmed using the red clump (RC) stars identified in the colour-magnitude diagrams. The absolute magnitudes of the RC stars have been the subject of many studies in the sense that they can be used as standard candles in distance estimation of the objects associated with them \citep[cf.][]{Alves00, Grocholski02, Groenewegen08, Yaz13, Bilir13a, Bilir13b}. \citet{Groenewegen08} calculated the mean absolute magnitude for the RC stars as $M_{K_s}=-1.54\pm0.04$ mag in the Two Micron Sky Survey \cite[2MASS;][]{Skrutskie06} photometric system, while the mean absolute magnitude in the $V$ band is given by \citet{Keenan99} as $0.7\leq M_V\leq1$ mag. 

In the colour-magnitude diagrams of NGC 6811, stars No: 2, 31, 57, 74 and 120 occupy the region of the RC stars, i.e. $0.7\leq(B-V)_0\leq1.2$ mag in the $V$ versus $(B-V)$ colour-magnitude diagram (Fig. 12). Additionally, four of these (31, 57, 74 and 120) lie in the giant region of the $V_0$ versus $J_0$ two-magnitude diagram in \citet{Bilir06} suggesting that they are evolved stars \citep[see also,][]{Bilir10}. We adopted the value $M_V=1\pm0.2$ mag as the absolute magnitude of the stars No: 2, 31, 57, 74 and 120, and used their apparent $V$ magnitudes to estimate their distances. Resulting values are given in Table 6. The estimated distances  for the cluster NGC 6811 from the four colour-magnitude diagrams (Table 5) are in agreement with the values inferred from the RC stars, within the errors. Thus, we confirm the distance of the cluster estimated by the procedure in situ, i.e. by fitting its colour-magnitude diagram to a standard main-sequence, via about a half dozen RC stars. 

\subsection{Metallicity Estimation by Means of Normalized UV-excess} 

We measured the metallicity of the cluster NGC 6811 by means of the normalized ultra-violet (UV) excesses, $\delta_{0.6}$, of 60 main-sequence cluster stars with spectral types F0-G0 and colours $0.3\leq (B-V)_0\leq 0.6$ mag. We evaluated the difference between the de-reddened $(U-B)_0$ colour index of a given star and the one corresponding to the Hyades star with the same de-reddened $(B-V)_0$ colour index, i.e. $\delta =(U-B)_{0,H}-(U-B)_{0,S}$, and normalized it to the UV-excess at $(B-V)_0=0.6$ mag, i.e. $\delta_{0.6}$, following the procedure in \citet{Karaali11}. Here, the subscripts $H$ and $S$ refer to ``Hyades'' and ``star'', respectively. The spectral types of the stars have been obtained from \citet{Cox00} by means of their de-reddened $(B-V)_0$ colour indices. 

Then, we plotted the histogram of normalized $\delta_{0.6}$ excesses for 60 main-sequence stars  with spectral types F0-G0 mentioned above and estimated the mode of this distribution. Finally, we transformed the mode to obtain the $[M/H]$ metallicity using the following equation in \citet{Karaali11}:
\begin{eqnarray}
[M/H]=-14.316(1.919)\delta_{0.6}^2-3.557(0.285)\delta_{0.6}\\ \nonumber
+0.105(0.039).
\end{eqnarray}

The two-colour diagram, $(U-B)_0$ versus $(B-V)_0$, of 60 main-sequence stars used in the calculation and their normalized UV-excess, $\delta_{0.6}$, histogram are given in Fig. 13. The mode for the $\delta_{0.6}$ distribution and the corresponding metallicity are $\delta_{0.6}=0.043\pm 0.004$ mag and $[M/H]=-0.074\pm0.020$ dex, respectively.   

\section{DISCUSSION}

In this paper, we present the results of our photometric observations and determine the structural and astrophysical parameters of the open cluster NGC 6811. We used the {\it Galaxia} model of \citet{Sharma11} and identified 133 likely field stars with $10\leq V\leq 18$ mag in the line-of-sight to the cluster. Thus, we could obtain a density profile for the cluster based only on its possible members, in addition to the density profile in situ, i.e. the density profile formed by the cluster and field stars. The measured core radius $r_c$ by the two procedures are in agreement with each other within their formal uncertainties. However, the central and background densities are different, as would be expected. \citet{Janes13} estimated nine sets of parameters for the density profile of the cluster NGC 6811. We compared our parameters, related to the cluster and field stars, with those of \citet{Janes13} found for the apparent magnitude interval $11<V<18$ (Table 7). As presented in Table 7, the core radius and the peak central stellar density parameters, found via the King (\citeyear{King62}) model are in very good agreement with each other, within the uncertainties of individual measurements. However, the background stellar density found here, $0.592\pm0.040$ stars/arcmin$^2$, is less than the one in \citet{Janes13}, $1.041\pm 0.048$ stars/arcmin$^2$. We should note that the amounts of the differences between the structural parameters in our study and the corresponding ones in \citet{Janes13} are at the level of the differences in \citet{Janes13} among different models, i.e. \citet{King62}, \citet{Plummer11}, and exponential model, for a given parameter.

We have two sets of colour excesses and distance moduli. The colour excess, distance moduli, the metallicity, and age of the cluster are all determined from four colour-magnitude diagram sets, simultaneously following a Bayesian approach. As a further test of the Bayesian approach employing independent methods, we determined the colour excesses, using two-colour diagrams, $U-B$ versus $B-V$, $V-I$ versus $V-R$ and $V-R$ versus $R-I$. The distance moduli have been estimated via colour-magnitude diagrams, $V$ versus $B-V$, $V$ versus $V-R$, $V$ versus $V-I$ and $V$ versus $R-I$. Finally, the metallicity of the cluster is measured using photometric UV-excesses of F-G type main-sequence stars. The colour excesses in the two methods are compatible for a given colour (Table 8). The same case holds also for the distance moduli (Table 5), as well, which further implies that the simultaneous estimation of the parameters is not affected by the degenaracies between the individual parameters.                    
               
The apparent distance modulus reported by \citet{Janes13}, $\mu=10.22\pm0.18$ mag, corresponds to a linear distance of $d=995\pm80$ pc. This is again compatible with our results, i.e. distances 
$d_1=960\pm70$ pc and $d_2=1000\pm50$ pc, estimated via the weighted mean distance moduli 10.06 and 10.15 mag in Table 5. Additionally, the distances of five RC stars in the range $990\leq d\leq 1127$ pc further confirm the distance estimation of the cluster NGC 6811, within the errors. The  age of the cluster NGC 6811 found in both studies, as $1.00\pm0.05$ and $1.00\pm0.17$ Gyr, here and \citet{Janes13}, respectively are in perfect agreement with each other.  The weighted mean of $E(B-V)$ colour excesses in the two sets, $0.05\pm0.01$ mag, is compatible with the one given by \citet{Janes13} as the best estimate for the $E(B-V)$ colour excess, $0.074\pm0.024$ mag, within the errors. Whereas, the best estimate for the metallicity in \citet{Janes13}, $[M/H]=-0.19\pm 0.03$ dex, is somewhat smaller than the weighted one in Table 5, $[M/H]=-0.10\pm0.01$ dex. We attribute this small difference to a possible degenarcy between all of these astrophysical parameters, i.e. slightly different colour excess, metallicity and distance values resulting in a similar age for the cluster. $E(B-V)$ colour excess and distance of the cluster found here have been confirmed by the two-colour and colour-magnitude diagrams, respectively. Hence, we conclude that the results in this study are not affected from the degeneracy in question.

We compared the $E(B-V)$ colour excess, mean distance, metallicity and age estimated via the colour-magnitude diagrams in our study with the corresponding ones appeared in the literature, summarized in Table 9. The $E(B-V)$ colour excesses estimated in the recent years, \citet{Janes13} and in this study, are almost half of the ones estimated formerly, \citet{Lindoff71}, \citet{Barkhatova78}, and \citet{Glushkova99}. While, there is a decrease in the distance estimation with increasing time. In these earlier studies, the metallicity of the cluster has been assumed as $[M/H]=0$ dex \citep{Lindoff71,Barkhatova78,Glushkova99}. However, the recent two studies show that the cluster is a bit metal-poor, i.e. $[M/H]=-0.19\pm0.03$ dex in \citet{Janes13} and $[M/H]=-0.10\pm0.01$ dex in our study. Still, there is a considerable difference between the two recent measurements. The metallicity of $[M/H]=-0.074\pm0.020$ dex estimated via the normalized UV-excess in our study is compatible with the value found following the Bayesian approach, as $[M/H]=-0.10\pm0.01$ dex. Hence, we conclude that NGC 6811 should be a bit metal-rich than the one claimed by \citet{Janes13}. There is a substantial difference between the ages estimated in the recent years and the ones appeared in the earlier studies, i.e. the cluster NGC 6811 is rather older than measured formerly. 

As a final remark, we would like to emphasize the use of {\it Galaxia} model of \citet{Sharma11} to separate the field stars from the likely members of the cluster. Due to its practicality, this procedure may lead to an accurate estimation of structural and astrophysical parameters for the clusters to be studied in the future.

\begin{acknowledgements}
We appreciate the referee for his/her valuable comments and suggestions. This work has been supported in part by the Scientific and Technological Research Council (T\"UB\.ITAK) 113F201 and 113F270. We thank to T\"UB\.ITAK for a partial support in using T100 telescope with project number 12BT100-324. EP acknowledge support by the SoMoPro II programme (3SGA5916). It was also supported by the grants GP14-26115P, 7AMB14AT015, the financial contributions of the Austrian Agency for International Cooperation in Education and Research (BG-03/2013 and CZ-09/2014). This research has made use of the WEBDA, SIMBAD, and NASA\rq s Astrophysics 
Data System Bibliographic Services. 
\end{acknowledgements}

\begin{table}
\caption{Log of observations, with exposure times
for each passband. $N$ denotes the number of exposure.} 
\begin{center}
\begin{tabular}{@{}cc@{}}
\hline \hline
Filter & Exp. time (s)$\times$N  \\
 \hline
$U$ & 90$\times$2, 60$\times$1, 360$\times$3  \\ 
$B$ & 20$\times$3, 90$\times$3 \\
$V$ & 5$\times$3, 60$\times$3  \\
$R$ & 2$\times$3, 20$\times$3  \\
$I$ & 1$\times$3, 10$\times$3  \\
\hline \hline
\end{tabular}\label{logobs}
\end{center}
\end{table}

\begin{table*}
\caption{Derived transformation and extinction coefficients. $k$ and $k'$ are primary and secondary extinction coefficients, respectively, while {\it $\alpha$} and $C$ are transformation coefficients.} 
\begin{center}
\begin{tabular}{ccccc}
\hline \hline
Band/Coefficient &       $k$ &     $k'$ & {\it $\alpha$} &       $C$ \\
\hline 
      $u$ & $0.597\pm0.032$ & $-0.058\pm0.032$ &        - &        - \\
      $b$ & $0.392\pm0.022$ & $-0.030\pm0.019$ & $0.930\pm0.043$ & $0.771\pm0.050$ \\
      {$v$} & $0.247\pm0.005^{(a)}$  &        - &        - &        - \\
            & $0.244\pm0.008^{(b)}$  &        - &        - &        - \\
            & $0.254\pm0.004^{(c)}$  &        - &        - &        - \\
      $r$ & $0.147\pm0.010$&        - & $-0.951\pm0.020$ & $0.921\pm0.028$ \\
      $i$ & $0.098\pm0.005$ &        - & $-1.038\pm0.008$ & $1.497\pm0.014$ \\
    $u-b$ &        - &        - & $0.929\pm0.054$  & $3.111\pm0.054$ \\
    $b-v$ &        - &        - & $0.056\pm0.006$  & $0.799\pm0.010$ \\
    $v-r$ &        - &        - & $0.108\pm0.006$  & $0.797\pm0.010$ \\
    $v-i$ &        - &        - & $0.052\pm0.006$  & $0.787\pm0.010$ \\
\hline \hline
\end{tabular}
\end{center}
{$^{a,b,c}$ calculated from $b-v$, $v-r$ and $v-i$, respectively.}
\end{table*}

\begin{table*}
\setlength{\tabcolsep}{2.2pt} 
\caption{Photometric catalogue for the open cluster NGC 6811. The complete table can be obtained electronically.} 
\begin{center}
\begin{tabular}{ccccccccccccccc}
\hline\hline
        ID &  $\alpha$ (J2000) &   $\delta$ (J2000) &      $V$ &   $U-B$ &   $B-V$ &  $V-R$ &   $V-I$ &   $R-I$ \\
\hline
    1 &  19  37 18.71 &  46  33  3.89 &    12.490$\pm$     0.002    &     0.115$\pm$     0.003    &     0.579$\pm$     0.002    &     0.365$\pm$     0.004    &     0.696$\pm$     0.005    &     0.329$\pm$     0.005   \\
    2 &  19  37 22.09 &  46  32 51.36 &    11.121$\pm$     0.001    &     0.703$\pm$     0.002    &     0.932$\pm$     0.001    &     0.515$\pm$     0.002    &     0.959$\pm$     0.002    &     0.440$\pm$     0.002   \\
    3 &  19  37 49.41 &  46  32 58.46 &    12.933$\pm$     0.003    &     0.172$\pm$     0.004    &     0.555$\pm$     0.003    &     0.332$\pm$     0.005    &     0.623$\pm$     0.007    &     0.290$\pm$     0.007   \\
    4 &  19  36 59.58 &  46  32 51.25 &    13.283$\pm$     0.003    &     1.168$\pm$     0.010    &     1.210$\pm$     0.004    &     0.679$\pm$     0.005    &     1.259$\pm$     0.006    &     0.573$\pm$     0.006   \\
    5 &  19  37 38.96 &  46  32 49.02 &    13.345$\pm$     0.003    &     0.009$\pm$     0.004    &     0.478$\pm$     0.004    &     0.324$\pm$     0.006    &     0.590$\pm$     0.008    &     0.263$\pm$     0.009   \\
         . & ... &  ... &     ... &      ... &      ... &      ... &      ... &      ... \\
         . & ... &  ... &     ... &      ... &      ... &      ... &      ... &      ... \\
         . & ... &  ... &     ... &      ... &      ... &      ... &      ... &      ... \\
\hline\hline
\end{tabular} 
\end{center}
\end{table*}

\begin{table*}
\caption{Mean photometric errors of the stars in the direction of the cluster NGC 6811. $N$ indicates the number of stars within the apparent magnitude range in the first column.} 
\begin{center}
\begin{tabular}{cccccccc}
\hline\hline
Magnitude Range &    $N$ &  $\sigma_V$ & $\sigma_{U-B}$ &  $\sigma_{B-V}$& $\sigma_{V-R}$ &  $\sigma_{V-I}$& $\sigma_{R-I}$\\
\hline
$10<V\leq12$ &         30 &      0.001 &      0.002 &      0.002 &      0.002 &      0.003 &      0.003 \\
$12<V\leq13$ &         45 &      0.002 &      0.003 &      0.003 &      0.004 &      0.006 &      0.006 \\
$13<V\leq14$ &         67 &      0.003 &      0.005 &      0.004 &      0.005 &      0.006 &      0.007 \\
$14<V\leq15$ &         85 &      0.002 &      0.004 &      0.002 &      0.003 &      0.004 &      0.004 \\
$15<V\leq16$ &        161 &      0.003 &      0.008 &      0.004 &      0.005 &      0.004 &      0.007 \\
$16<V\leq17$ &        289 &      0.005 &      0.015 &      0.008 &      0.008 &      0.011 &      0.012 \\
$17<V\leq18$ &        394 &      0.009 &      0.034 &      0.014 &      0.015 &      0.020 &      0.022 \\
$18<V\leq19$ &        378 &      0.015 &      0.069 &      0.026 &      0.027 &      0.037 &      0.040 \\
$19<V\leq20$ &        142 &      0.025 &      0.095 &      0.040 &      0.046 &      0.070 &      0.076 \\
\hline\hline
\end{tabular}  
\end{center}
\end{table*}

\begin{table*}
\setlength{\tabcolsep}{3pt} 
\caption{Colour excesses, distance moduli ($\mu_1$), metallicities and ages estimated simultaneously using four colour-magnitude diagrams. $\mu_2$ indicates the distance moduli determined individually, and $d$ denotes the linear distances calculated using $\mu_2$.} 
\begin{center}
\begin{tabular}{ccccccc}
\hline\hline
Colour-Magnitude & Colour Excess & $\mu_1$ & $[M/H]$ & Age   & $\mu_2$ & $d$\\
   Diagram       &      (mag)    & (mag) &   (dex) & (Gyr)  & (mag) & (pc)\\
\hline
$V$ versus $B-V$ & $E(B-V)=0.05\pm0.01$ & $10.20\pm0.18$ & $-0.10\pm0.01$ & $1.00\pm0.05$ & $10.22\pm0.15$ & $1030\pm71$\\
$V$ versus $V-R$ & $E(V-R)=0.05\pm0.01$ & $10.05\pm0.15$ & $-0.10\pm0.01$ & $1.00\pm0.05$ & $10.26\pm0.25$ & $1010\pm122$\\
$V$ versus $V-I$ & $E(V-I)=0.08\pm0.01$ & $ 9.95\pm0.17$ & $-0.10\pm0.01$ & $1.00\pm0.05$ & $10.22\pm0.25$ & $1010\pm119$\\
$V$ versus $R-I$ & $E(R-I)=0.04\pm0.01$ & $10.05\pm0.16$ & $-0.10\pm0.01$ & $1.00\pm0.05$ & $10.32\pm0.46$ & $1054\pm230$\\
\hline
\multicolumn{2}{r}{Mean} & $10.06\pm0.08$ & $-0.10\pm0.01$ & $1.00\pm0.05$ & $10.23\pm0.11$ & \\
\hline\hline
\end{tabular}    
\end{center}
\end{table*}

\pagebreak
\pagebreak
\pagebreak

\begin{table*}
\caption{Equatorial coordinates, apparent $V$ (in this study) and $J$ \citep{Cutri03} magnitudes, de-reddened $V_0$ and $J_0$ magnitudes, $B-V$ colours and the de-reddened ones $(B-V)_0$, and distances for five RC star candidates in NGC 6811. Numbers in parenthesis are errors.} 
\begin{center}
\begin{tabular}{cccccccccc}
\hline\hline
ID  & $\alpha$ (J2000)& $\delta$ (J2000) &    $V$     &      $B-V$ &        $J$ &      $V_0$ &  $(B-V)_0$ & $J_0$ & $d$ \\
    & (hh:mm:ss)      & (dd:mm:ss)       &    (mag)   &      (mag) &      (mag) &      (mag) &      (mag) & (mag) & (pc) \\
\hline
 2  & 19:37:22.09     & 46:32:51.36      &  11.121(1) &   0.932(1) &  9.489(20) &     10.978 &      0.886 & 9.448 & $ 990\pm87$ \\
31  & 19:36:55.83     & 46:27:38.25      &  11.354(1) &   0.932(2) &  9.671(20) &     11.211 &      0.886 & 9.630 & $1102\pm92$ \\
57  & 19:37:02.70     & 46:23:13.76      &  11.319(1) &   0.949(2) &  9.599(20) &     11.176 &      0.903 & 9.558 & $1084\pm95$ \\
74  & 19:38:04.90     & 46:24:42.61      &  11.403(1) &   0.920(2) &  9.714(20) &     11.260 &      0.874 & 9.673 & $1127\pm97$ \\
120 & 19:36:57.15     & 46:22:43.22      &  11.247(1) &   0.928(2) &  9.541(20) &     11.104 &      0.882 & 9.500 & $1049\pm99$ \\
\hline\hline
\end{tabular}  
\end{center}
\end{table*}

\begin{table*}
\caption{Structural parameters of NGC 6811 found in our study and in \citet{Janes13} using the King model.} 
\begin{center}
\begin{tabular}{|c|cc|c|}
\hline\hline
           & \multicolumn{2}{|c|}{This study}     & \multicolumn{1}{|c|}{Janes et al. (2013)} \\  
\hline
 Parameter & Cluster & Cluster $+$              &  King Model \\
           &  stars  & field stars              &  \\
\hline
     $f_0$ &  $0.921\pm0.115$ & $1.566\pm0.085$ &  $1.640\pm0.482$ \\
   $f_{bg}$&  $0.438\pm0.048$ & $0.592\pm0.275$ &  $1.041\pm0.048$ \\
     $r_c$ &  $3.600\pm0.563$ & $3.860\pm0.275$ &  $4.404\pm1.388$ \\
\hline\hline
\end{tabular}  
\end{center}
\end{table*}

\begin{table*}
\caption{Colour excesses estimated using the two-colour diagrams (TCD) and colour-magnitude diagrams (CMD) of the cluster.} 
\begin{center}
\begin{tabular}{ccc}
\hline\hline
Colour excess&        TCD       &        CMD    \\
\hline
    $E(U-B)$ &  $0.033\pm0.011$ &         $-$  \\
    $E(B-V)$ &  $0.046\pm0.012$ & $0.05\pm0.01$ \\
    $E(V-R)$ &  $0.053\pm0.010$ & $0.05\pm0.01$ \\
    $E(V-I)$ &  $0.102\pm0.018$ & $0.08\pm0.01$ \\
    $E(R-I)$ &  $0.046\pm0.010$ & $0.04\pm0.01$ \\
\hline\hline
\end{tabular}  
\end{center}
\end{table*}  

\begin{table*}
\caption{Astrophysical parameters of NGC 6811 in this study and those collected from the literature.} 
\begin{center}
\begin{tabular}{cccccc}
\hline\hline
 ID &              Study       &    $E(B-V)$     &   $d$        &  $[M/H]$       &  $Age$       \\
    &                          &    (mag)        &   (pc)       &  (dex)         &  (Myr)       \\
\hline
  1 & Lindoff (1972)           &       0.16      &    1100      &          0     &        500   \\
  2 & Barkhatova et al. (1978) &       0.14      & $1150\pm110$ &          0     &        800   \\
  3 & Glushkova et al. (1999)  & $0.12\pm0.02$   & $1040\pm45 $ &          0     & $700\pm100 $ \\
  4 & Janes et al. (2013)      & $0.074\pm0.024$ & $1000\pm50 $ & $-0.19\pm0.03$ & $1000\pm170$ \\
  5 & This study (2014)        & $0.050\pm0.010$ & $960\pm70  $ & $-0.10\pm0.01$ & $1000\pm50$ \\
\hline\hline
\end{tabular}  
\end{center}
\end{table*}  

\begin{figure*}
\begin{center}
\includegraphics[scale=0.60, angle=0]{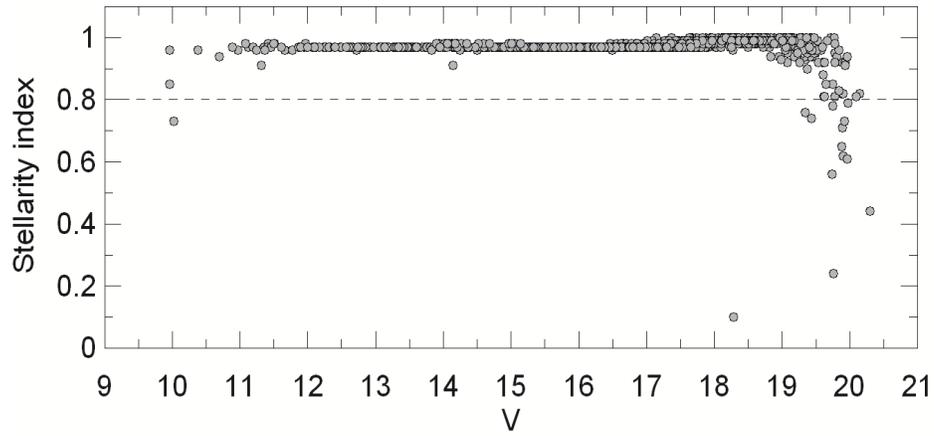}
\caption[] {$V$ magnitude versus stellarity-index diagram for 1605 objects observed in the direction to the cluster NGC 6811. Dashed line corresponds to the stellarity (SI) index of 0.8. 14 objects with $SI < 0.8$ are classified as non-stellar objects.} 
\end{center}
\end {figure*} 
\pagebreak

\begin{figure*}
\begin{center}
\includegraphics[scale=0.60, angle=0]{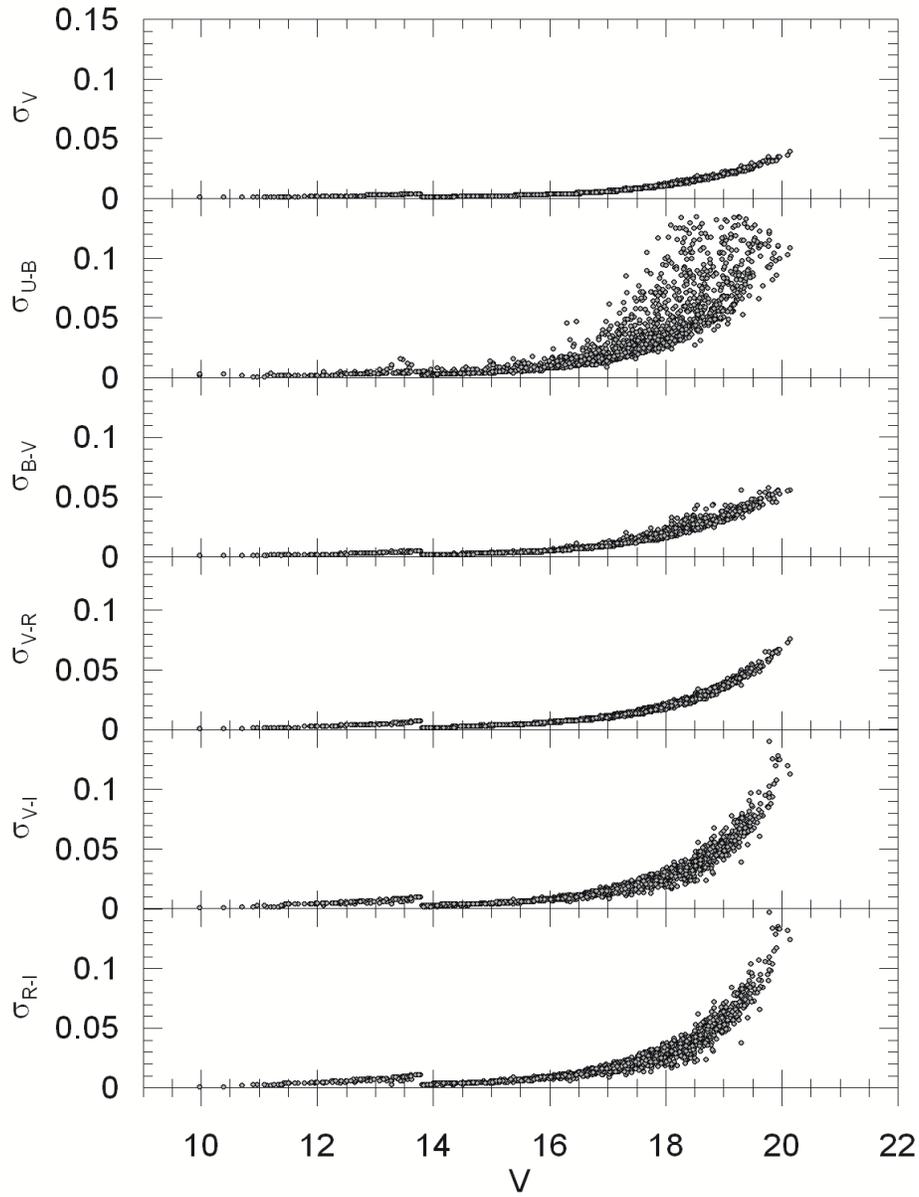}
\caption[] {Colour and magnitude errors of the stars observed in the line-of-sight of the cluster NGC 6811, as a function of $V$ apparent magnitude.} 
\end{center}
\end {figure*} 

\begin{figure*}
\begin{center}
\includegraphics[scale=0.80, angle=0]{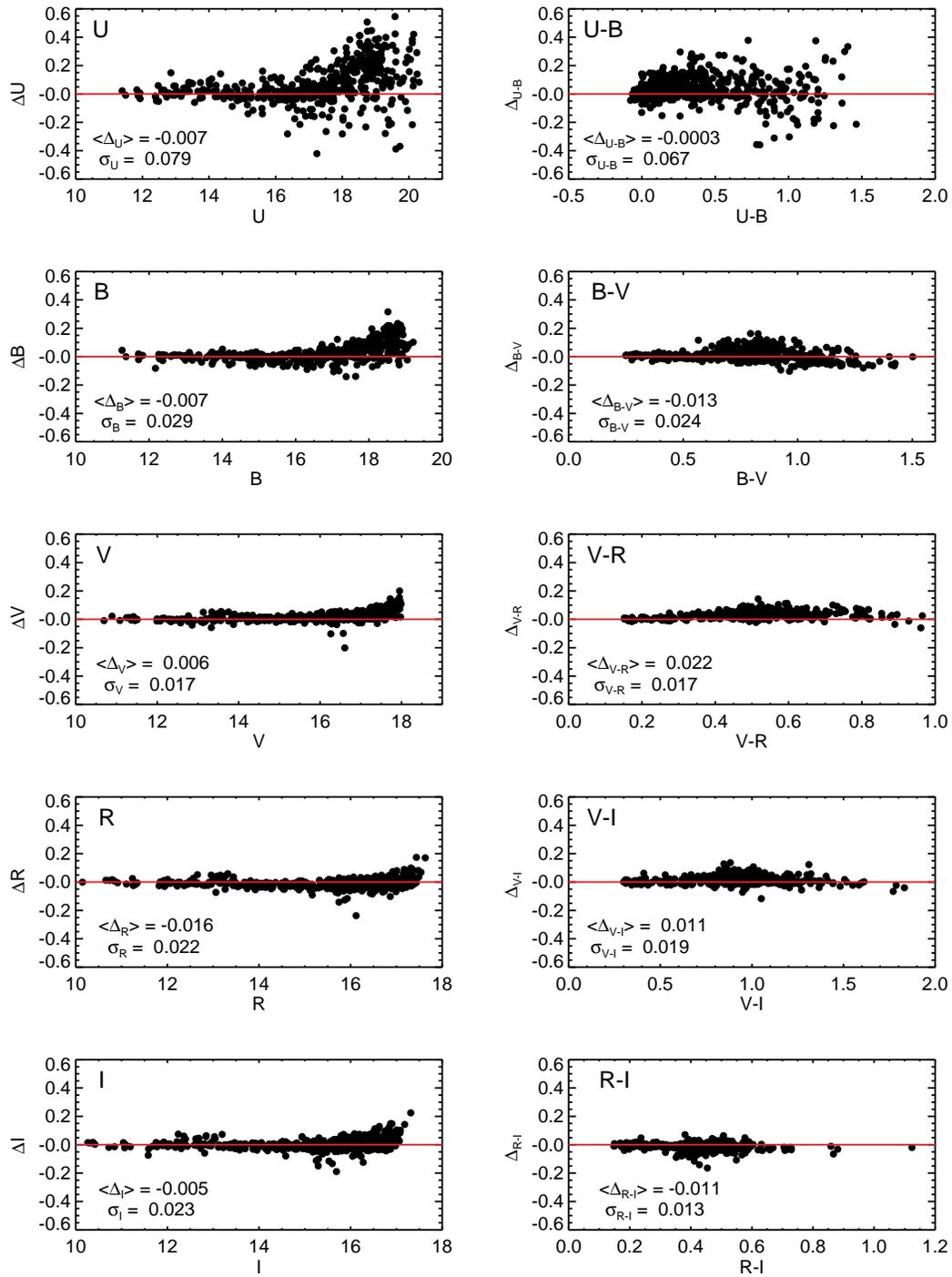}
\caption[] {Comparison of the magnitudes and colours for a sample of stars observed in our study and in \citet{Janes13}. The means and standard deviations of the differences are shown in panels.} 
\end{center}
\end {figure*} 

\begin{figure*}
\begin{center}
\includegraphics[scale=0.70, angle=0]{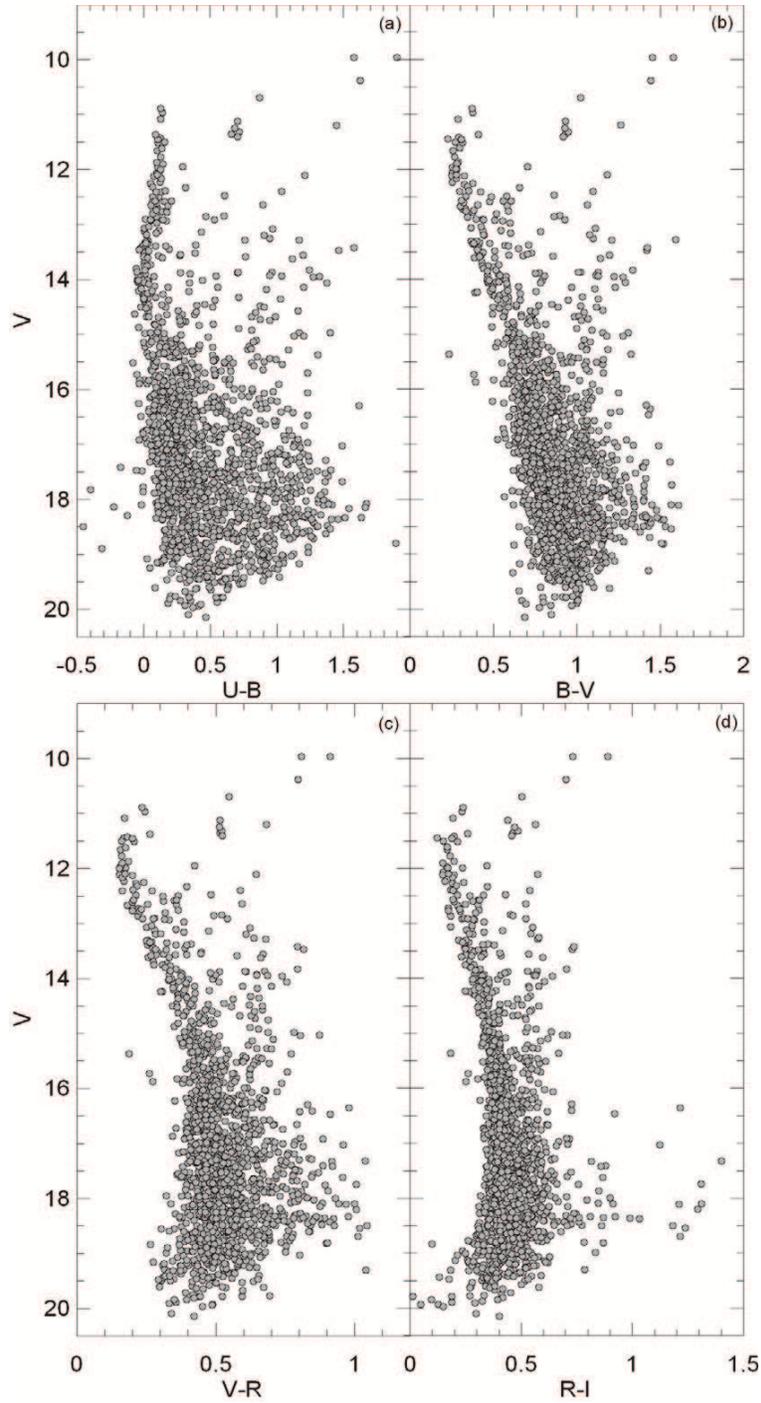}
\caption[] {Colour-magnitude diagrams for the cluster NGC 6811. (a) $V$ versus $U-B$, (b) $V$ versus $B-V$, (c) $V$ versus $ V-R$, and (d) $V$ versus $R-I$.}      
\end{center}
\end {figure*} 

\begin{figure*}
\begin{center}
\includegraphics[scale=0.80, angle=0]{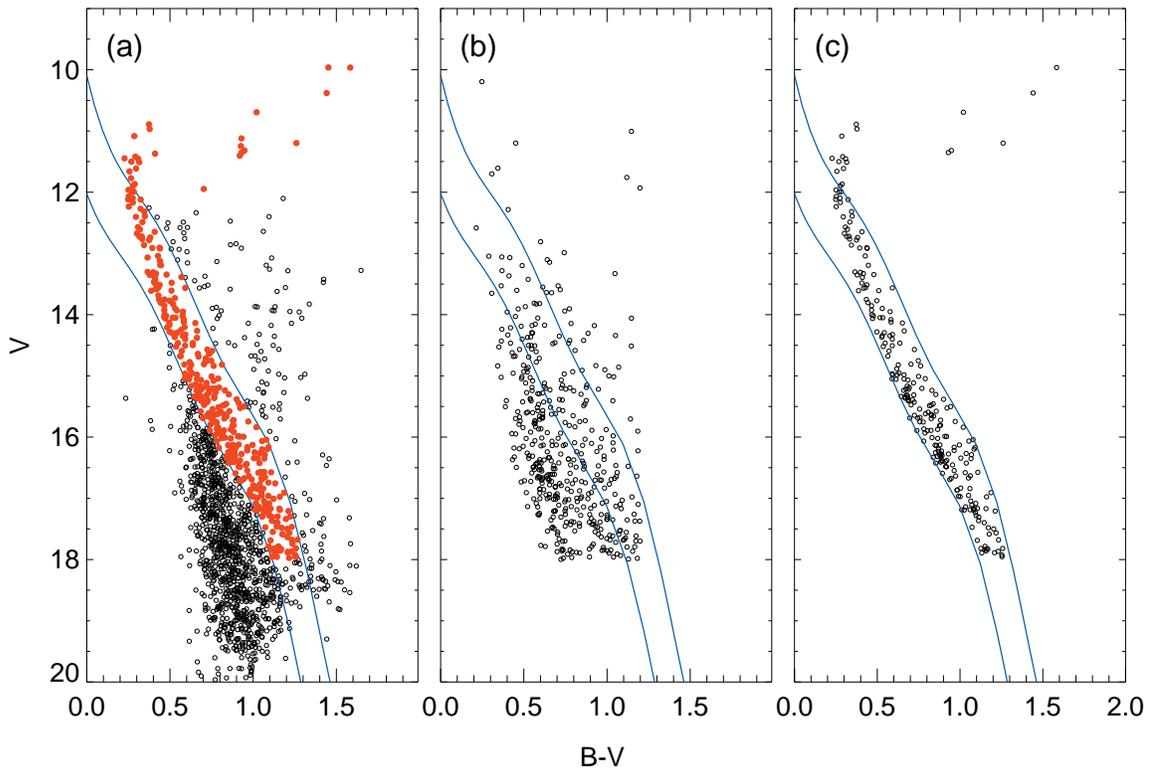}
\caption[] {$V\times B-V$ colour-magnitude diagrams for three samples of stars: (a) For 1591 stars with apparent magnitude $10\leq V\leq 20$, where the lines indicate the zero age and zero metallicity Padova isochrone and the one moved by an amount of 0.75 mag to the bright $V$ magnitudes. The number of stars shown with red circles is 387. (b) For 446 synthetic stars whose $V$ and $B-V$ data are estimated by the {\it Galaxia} model, where 133 of them lie within the isochrones defined in panel (a). (c) Resulting 254 likely cluster member stars with $10\leq V\leq 18$ mag remained after the subtraction of the randomly selected stars.}      
\end{center}
\end {figure*}

\begin{figure*}
\begin{center}
\includegraphics[scale=0.60, angle=0]{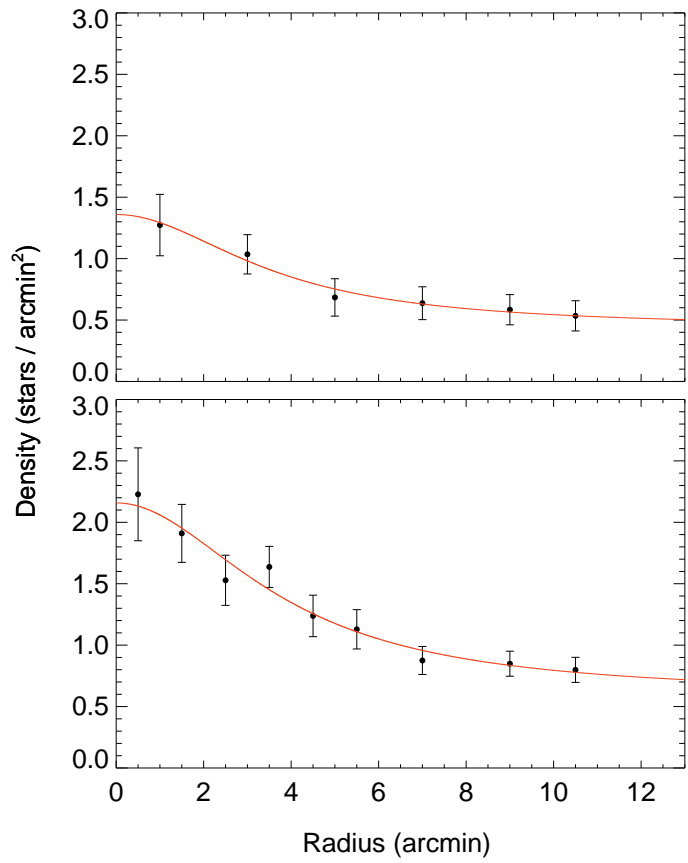}
\caption[] {Stellar density profile of the cluster NGC 6811 for two sets of main-sequence stars with $10\leq V\leq 18$ mag. Upper Panel: For cluster stars, and lower panel: for cluster plus field stars.}      
\end{center}
\end {figure*}

\begin{figure*}
\begin{center}
\includegraphics[scale=0.60, angle=0]{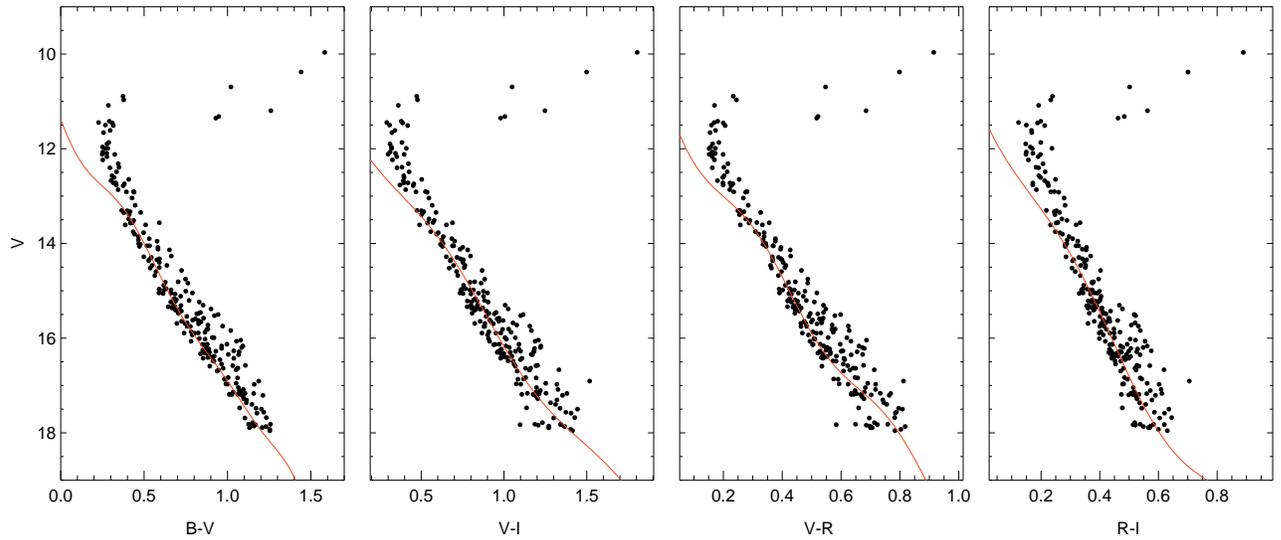}
\caption[] {Colour-magnitude diagrams used for the distance modulus estimation of the cluster. The line in the $V$ versus $B-V$ diagram indicates the ZAMS of \citet{Sung13}, while the ones in the diagrams $V$ versus $V-R$, $V$ versus $V-I$ and $V$ versus $R-I$ are adopted from ZAMS of Padova synthetic stellar library \citep{Marigo08}.}      
\end{center}
\end {figure*}

\begin{figure*}
\begin{center}
\includegraphics[scale=0.40, angle=0]{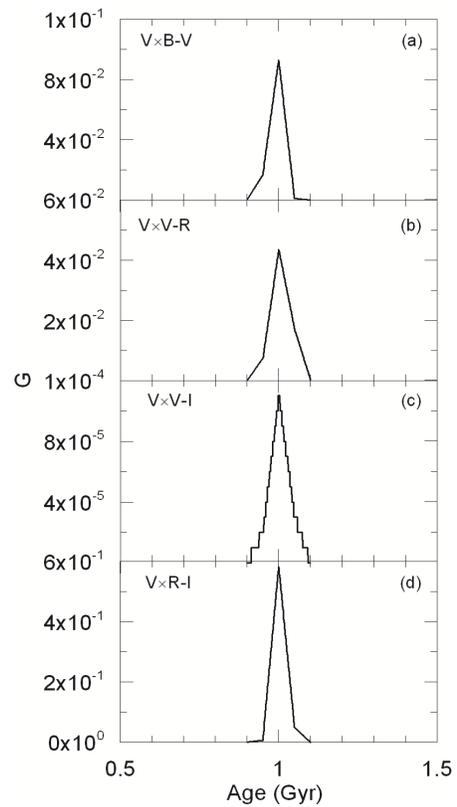}
\caption[] {Distribution of $G(\tau)$ parameter in terms of age for four colour-magnitude diagrams.}      
\end{center}
\end {figure*}

\begin{figure*}
\begin{center}
\includegraphics[scale=0.60, angle=0]{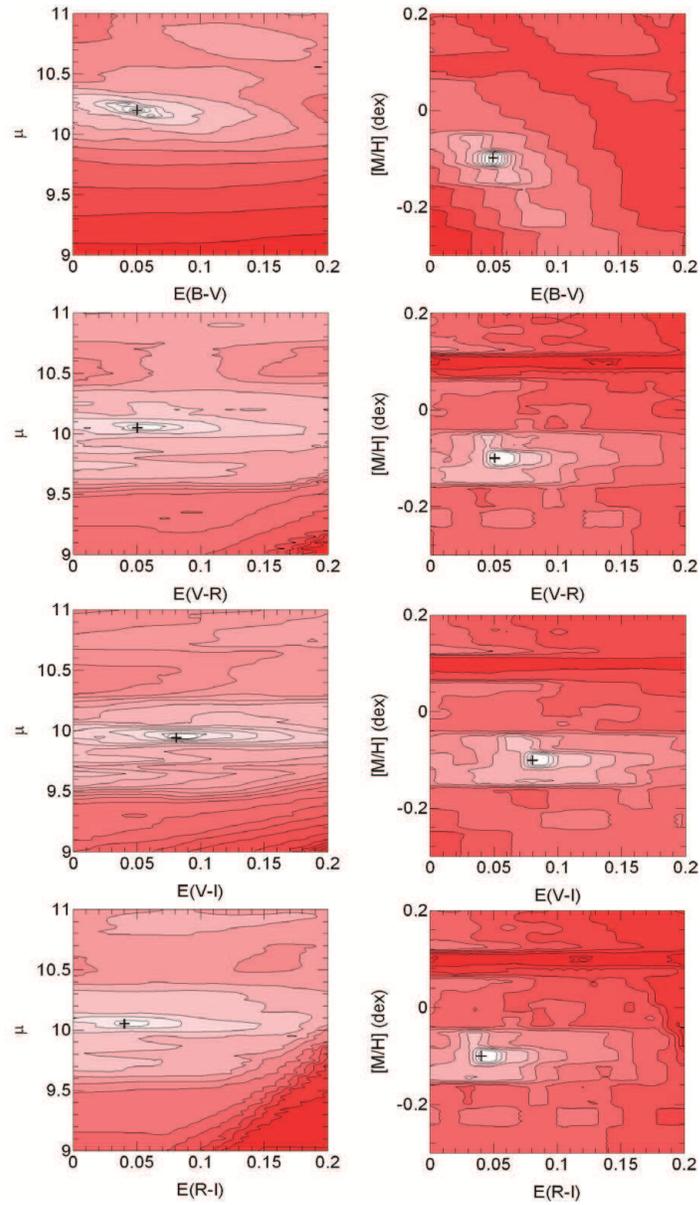}
\caption[] {Contour diagrams for distance modulus, metallicity and colour excess, for different colours used for estimation of the age, distance modulus, metallicity and colour excess of the cluster simultaneously. The symbol ``+'' indicates the most probable value of the corresponding parameter.}
\end{center}      
\end {figure*}

\begin{figure*}
\begin{center}
\includegraphics[scale=0.40, angle=0]{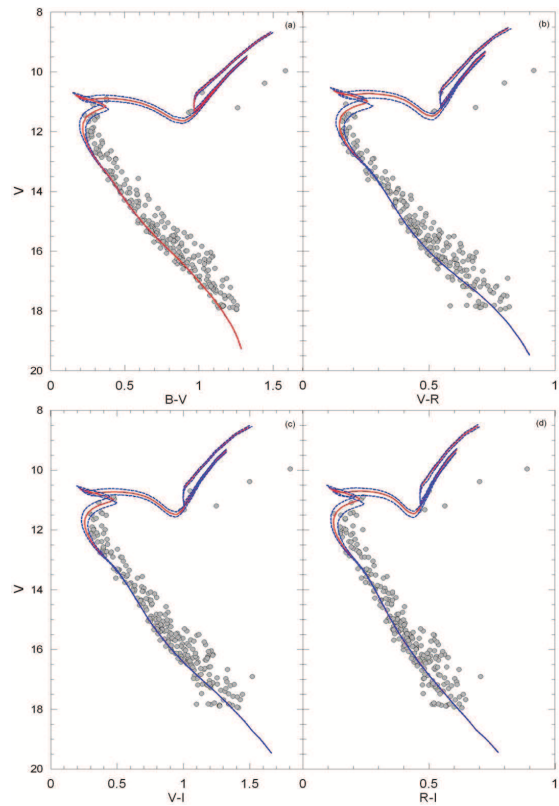}
\caption[] {Four colour-magnitude diagrams, $V$ versus $B-V$, $V$ versus $V-R$, $V$ versus $V-I$ and $V$ versus $R-I$, for the cluster stars with $10\leq V\leq 18$ mag, fitted to the isochrone determined in this study (red line). The dotted blue lines indicate the isochrones with estimated age plus its error.}
\end{center}      
\end {figure*}

\begin{figure*}
\begin{center}
\includegraphics[scale=0.80, angle=0]{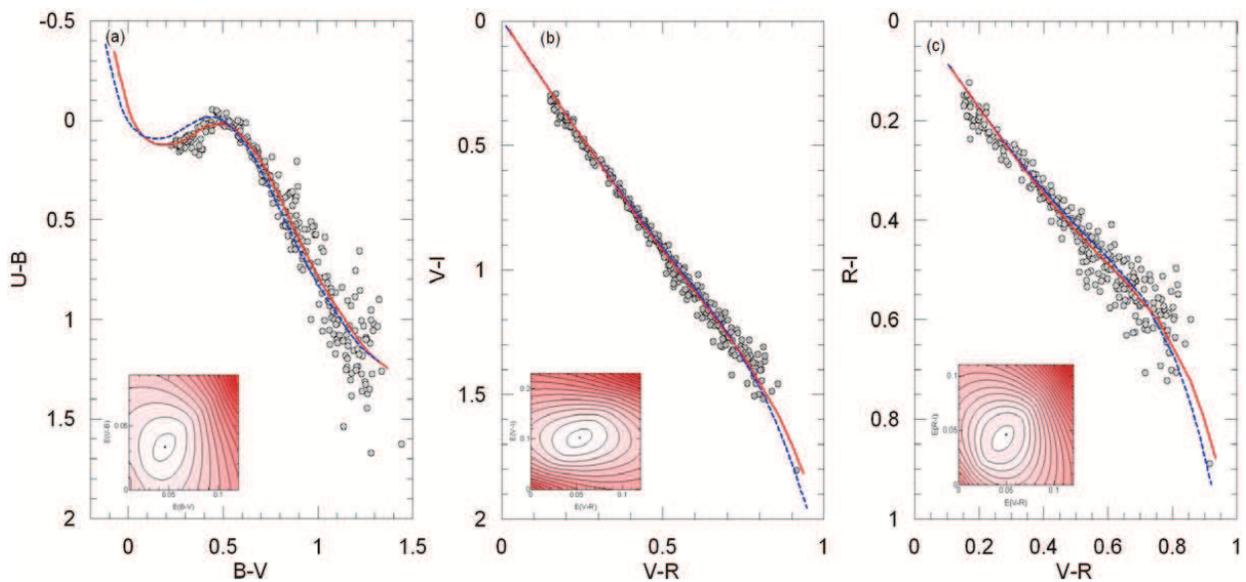}
\caption[] {Two-colour diagrams for the main-sequence stars with $12\leq V\leq 18$ mag of NGC 6811. (a) The standard main-sequence of \citet{Sung13} for $U-B$ versus $B-V$, (b) $V-I$ versus $V-R$ and (c) $V-R$ versus $R-I$, both adopted from the Padova synthetic stellar library \citep{Marigo08}. In each panel, the reddened and de-reddened main-sequence curves fitted to the cluster stars are indicated as solid red and dotted blue lines, respectively. Also, the contours of $\chi^2$ distributions are also shown in the left corner of each diagram.}      
\end{center}
\end {figure*}

\begin{figure*}
\begin{center}
\includegraphics[scale=0.70, angle=0]{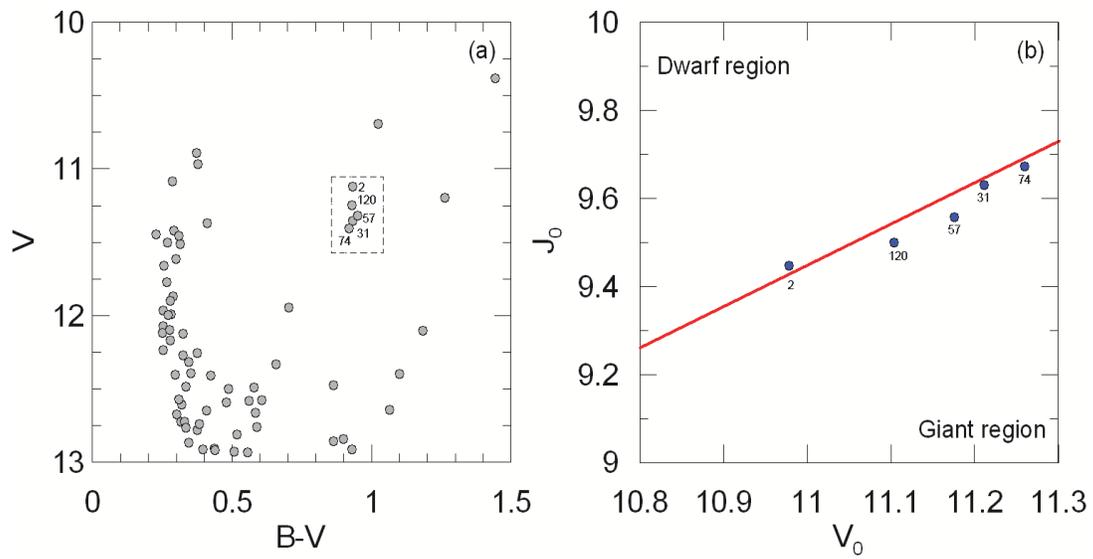}
\caption[] {Colour-magnitude diagram of the RC stars (a) and their position in the two-magnitude diagram, $J_0$ versus $V_0$ (b). The line, $J_0=0.957\times V_0-1.079$, which separates the stars into dwarf and giant categories is adopted from \citet{Bilir06}. The star No: 2 is also adopted as giant (RC star) due to its proximity to the line.}
\end{center}      
\end {figure*}

\begin{figure*}
\begin{center}
\includegraphics[scale=0.50, angle=0]{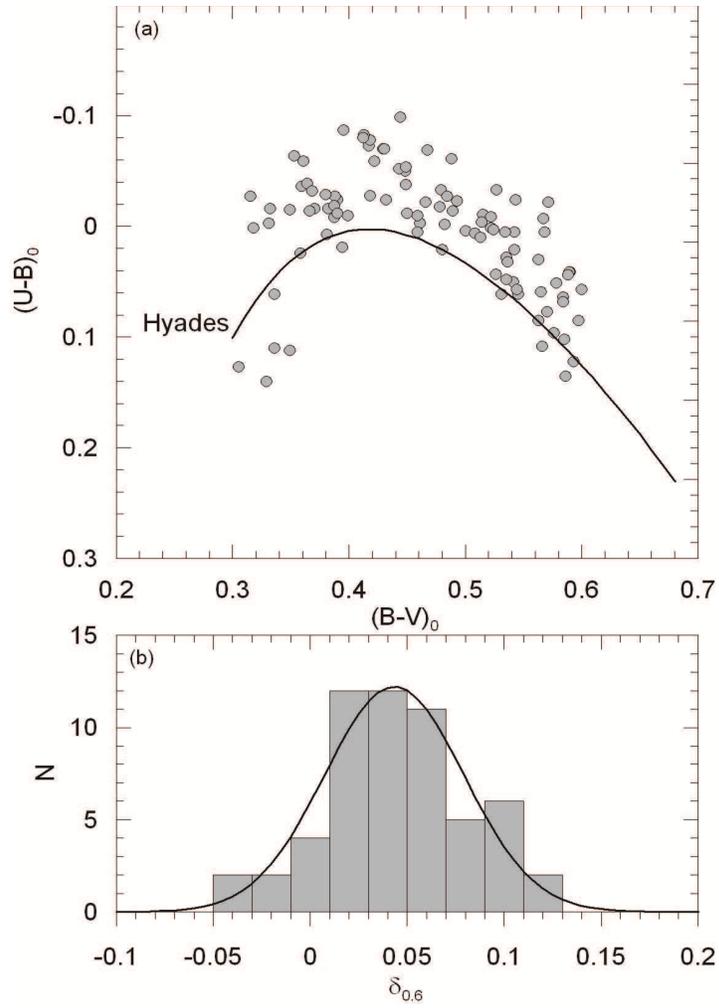}
\caption[] {Two-colour diagram (a) and the histogram for the reduced UV-excess (b) for 60 main-sequence stars used for the metallicity estimation of NGC 6811.}
\end{center}      
\end {figure*} 

\end{document}